\documentclass{PoS}

\usepackage{graphicx}
\usepackage{dcolumn}
\usepackage{bm}
\usepackage{epsfig}

\newcommand{\be}{\begin{eqnarray}}
\newcommand{\ee}{\end{eqnarray}}
\newcommand{\nn}{\nonumber}
\newcommand{\gev}{{\rm GeV}}
\newcommand{\mev}{{\rm MeV}}
\newcommand{\<}{\langle}
\renewcommand{\>}{\rangle}

\title{Vector and scalar form factors for K- and D-meson semileptonic decays from twisted mass fermions with $N_f = 2$}

\ShortTitle{Semileptonic Form Factors from Two-Flavor Lattice QCD}

\author{ETM Collaboration}

\author{S.~Di Vita, V.~Lubicz, C.~Tarantino\\
Dipartimento di Fisica, Universit{\`a} di Roma Tre, Via della Vasca Navale 84, I-00146 Roma, Italy\\
INFN, Sezione di Roma Tre, Via della Vasca Navale 84, I-00146 Roma, Italy}

\author{B.~Haas\\
LPT, Universit\'e Paris Sud, Centre d'Orsay, 91405 Orsay-Cedex, France}

\author{F.~Mescia\\
Dep.~ECM and ICC, Universitat de Barcelona, Diagonal 647, 08028 Barcelona, Spain}

\author{\speaker{S.~Simula}\\
INFN, Sezione di Roma Tre, Via della Vasca Navale 84, I-00146 Roma, Italy}

\abstract{We present lattice results for the form factors relevant in the $K \to \pi \ell \nu_\ell$ 
and $D \to \pi \ell \nu_\ell$ semileptonic decays, obtained from simulations with two flavors 
of dynamical twisted-mass fermions and pion masses as light as $260~\mev$.
For $K \to \pi \ell \nu$ decays we discuss  the estimates of the main sources of systematic 
uncertainties, including the quenching of the strange quark, leading to our final result 
$f_+(0) = 0.9560(57)_{\mbox{stat.}}(62)_{\mbox{syst.}}$. 
Combined with the latest experimental data, our value of $f_+(0)$ implies for the CKM 
matrix element $|V_{us}|$ the value $0.2267(5)_{\mbox{exp.}}(20)_{f_+(0)}$ consistent 
with the first-row CKM unitarity.
For $D \to \pi \ell \nu_\ell$ decays the application of Heavy Meson Chiral Perturbation 
Theory allows to extrapolate our results for both the scalar and the vector form factors at 
the physical point with quite good accuracy, obtaining a nice agreement with the 
experimental data.
In particular at zero-momentum transfer we obtain $f_+(0) = 0.64(5)$.
A preliminary analysis of the discretization effects is presented and discussed.
}

\FullConference{The XXVII International Symposium on Lattice Field Theory - LAT2009\\
		           July 26-31 2009\\
		           Peking University, Beijing, China}

\begin{document}

\section{Introduction}

Semileptonic decays of pseudoscalar mesons can provide important information on the 
weak mixing of quark flavors, which in the Standard Model constructs the well-known 
CKM matrix \cite{CKM}.
In order to extract from the experimental data precise values of the relevant CKM entries 
it is necessary to determine precisely the matrix elements of the weak hadronic current.

In the case of the semileptonic decay $H \to P \ell \nu_\ell$ the matrix element of the 
weak vector current can be written in terms of two form factors, the vector, $f_+(q^2)$, 
and the scalar, $f_0(q^2)$ ones, namely
 \be
    \< P(p_P) | V^\mu | H(p_H) \> = (p_P + p_H - \Delta)^\mu ~ f_+(q^2) + \Delta^\mu ~ f_0(q^2) ~ ,
    \label{eq:HP}
 \ee
where $\Delta \equiv q ~ (M_H^2 - M_P^2) / q^2$ and  $q \equiv p_H - p_P$ is the 4-momentum 
transfer.

In this contribution we present the lattice results for the vector and scalar form factors obtained 
by the European Twisted Mass (ETM) Collaboration in the case of the $K \to \pi \ell \nu_\ell$ and 
$D \to \pi \ell \nu_\ell$ semileptonic decays, which are relevant for the determination of the CKM 
matrix elements $|V_{us}|$ (known also as the Cabibbo's angle) and $|V_{cd}|$, respectively . 

\section{$K \to \pi \ell \nu$ decays}

The relevant hadronic quantity in the case of the $K \to \pi \ell \nu$ decays is the vector form 
factor at zero-momentum transfer, $f_+(0)$.
Its first determination dates back to the eighties, i.e.~to the work of Ref.~\cite{LR}, in which 
Chiral Perturbation Theory (ChPT) and the quark model were employed.

The determination of $f_+(0)$ using lattice QCD started only more recently with the quenched 
calculation of Ref.~\cite{SPQCDR}, where it was shown how $f_+(0)$ can be determined 
at the physical point with $\simeq 1 \%$ accuracy.
The findings of Ref.~\cite{SPQCDR} triggered various unquenched calculations of $f_+(0)$, 
namely those of Refs.~\cite{JLQCD,RBC06,QCDSF} with $N_f = 2$ and pion masses 
above $\simeq 500~\mev$ and the recent one of Ref.~\cite{RBC08} with $N_f = 2 + 1$ 
and pion masses starting from $330~\mev$.

In Ref.~\cite{ETMC09} a new lattice result for $f_+(0)$, namely
 \be
    f_+(0) = 0.9560 \pm 0.0057_{\mbox{stat.}} \pm 0.0062_{\mbox{syst.}} = 0.9560 \pm 0.0084 ~ ,
    \label{eq:final}
 \ee
was obtained by the ETM Collaboration using gauge configurations with $N_f = 2$ flavors of dynamical 
twisted-mass quarks \cite{ETMC} and simulating pion masses from $260~\mev$ up to $575~\mev$.

Our new determination (\ref{eq:final}) agrees very well with the Leutwyler-Roos result \cite{LR} and 
with previous lattice calculations at $N_f = 0$ \cite{SPQCDR}, $N_f = 2$ \cite{JLQCD,RBC06,QCDSF} 
and $N_f = 2 + 1$ \cite{RBC08}.
Using the latest experimental determination of the product $|V_{us}| f_+(0) = 0.21668(45)$ 
\cite{PDG,FlaviaNet} we get from (\ref{eq:final})
 \be
    |V_{us}| = 0.2267 \pm 0.0005_{\mbox{exp.}} \pm 0.0020_{f_+(0)} ~ .
    \label{eq:Vus}
 \ee
Combining this value with $|V_{ud}| = 0.97418(27)$ and $|V_{ub}| = 0.00393(36)$ from PDG2008 
\cite{PDG} the first-row CKM unitarity relation becomes
 \be
    |V_{ud}|^2 + |V_{us}|^2 + |V_{ub}|^2 = 1.0004 \pm 0.0015 .
    \label{eq:unitarity}
 \ee

Our final value (\ref{eq:final}) includes the estimates of all sources of systematic errors: discretization, 
finite size effects (FSE's), $q^2$-dependence, chiral extrapolation and the effects of quenching the 
strange quark.
In Ref.~\cite{ETMC09} the chiral extrapolation and the related uncertainty on $f_+(0)$ were investigated 
using both SU(3) and, for the first time, SU(2) ChPT \cite{SU2}, obtaining fully consistent results.
Also the $q^2$-dependence of the form factors was investigated by considering different functional 
forms.
The systematic error associated both to the $q^2$-dependence and to the chiral extrapolation was
determined quite accurately and turned out to be 0.0035 \cite{ETMC09}.

We illustrate now in more details the estimates of the remaining sources of systematic errors, namely 
finite size, discretization and the quenching of the strange quark.

{\it Finite Size.} We have performed simulations close to  $M_\pi \simeq 300~\mev$ using two lattice 
volumes, $24^3 \cdot 48 ~ a^4$ and $32^3 \cdot 64~ a^4$, for a lattice spacing equal to $a \simeq 
0.088$ fm. 
The two simulations correspond to $M_\pi L \simeq 3.2$ and $4.2$, respectively.
As described in Ref.~\cite{ETMC09}, a smooth interpolation of $f_+(0)$ at the physical strange quark 
mass can be obtained by fixing the combination ($2 M_K^2 - M_\pi^2$) at its physical value.
Thus for each pion mass $M_\pi$ a reference kaon mass $M_K^{ref}$ is defined as
 \be 
 2 [M_K^{ref}]^2 - M_\pi^2 = 2 [M_K^{phys}]^2 - [M_\pi^{phys}]^2 
 \label{eq:MKref}
 \ee
with $M_\pi^{phys} = 135.0~\mev$ and $M_K^{phys} = 494.4~\mev$.

\begin{figure}[!hbt]

\centerline{\includegraphics[scale=0.8]{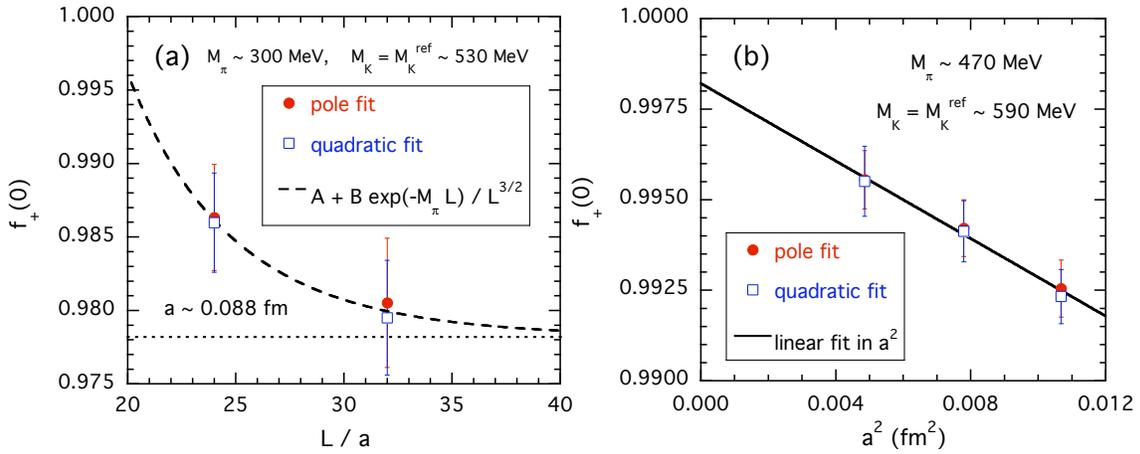}}

\caption{\it Vector form factor at zero-momentum transfer, $f_+(0)$, versus the lattice size in lattice 
units (a) and the squared lattice spacing (b). The values of the pion mass are reported in the inset, 
while the kaon mass is fixed at the corresponding reference values given by 
Eq.~(\protect\ref{eq:MKref}).
In (a) the dotted line represents the value $f_+(0) = A$ in the limit of infinite volume.}

\label{fig:fplus}

\end{figure}

The results for $f_+(0)$, obtained adopting either the pole-dominance or a quadratic fit for 
describing the $q^2$-dependence of the form factors (see Ref.~\cite{ETMC09}), are shown 
in Fig.~\ref{fig:fplus}(a) versus the lattice size $L / a$.
For matrix elements like $ \< \pi | V^\mu | K \>$, involving one particle in the final states, FSE are 
known to be exponentially suppressed.
Assuming a dependence of the form $A + B e^{-M_\pi L} / L^{3/2}$ the residual FSE, 
corresponding to the difference between the value at infinite volume and the one 
calculated at the largest lattice volume, turns out to be equal to $0.0018$.

{\it Discretization.} We have performed simulations at $M_\pi \simeq 470~\mev$ using 
three lattice spacings: $a \simeq 0.069, 0.088$ and $0.103~\mbox{fm}$.
The results for $f_+(0)$, shown in Fig.~\ref{fig:fplus}(b), exhibits a clear, linear (in $a^2$) 
increase toward the continuum limit, consistent with the automatic 
${\cal{O}}(a)$-improvement at maximal twist \cite{improvement}.
The difference between the value in the continuum limit and the one at $a \simeq 0.088$ fm 
is  equal to $0.0037$, which represents our estimate of the contribution of discretization 
effects to the systematic error in Eq.~(\ref{eq:final}).
A complete study of the scaling property of $f_+(0)$ at various pion masses is in progress.
It  will allow us to compute the continuum limit, reducing in this way significantly the error due 
to lattice artifacts.

{\it Quenching of the strange quark.} The effect of our partially quenched (PQ) setup can be 
estimated within SU(3) ChPT, which provides a systematic expansion of $f_+(0)$ of the type 
 \be
      f_+(0) = 1 + f_2 + f_4 + f_6 + ... ~ , 
      \label{eq:SU3}
  \ee
where $f_n = {\cal{O}}[M_{K,\pi}^n / (4 \pi f_\pi)^n]$ and the first term is equal to unity due to 
the vector current conservation in the SU(3) limit.

Because of the Ademollo-Gatto theorem \cite{AG}, valid also in both quenched (Q) 
\cite{SPQCDR} and PQ \cite{PQ}  setups, the first correction $f_2$ does not receive 
contributions from the local operators of the effective theory and can be computed 
unambiguously (for any $N_f$) in terms of the kaon and pion masses and the 
pion decay constant $f_\pi$. 
At the physical point it takes the values: $f_2^Q = +0.022$ in the quenched case $N_f = 0$ 
\cite{SPQCDR}, $f_2^{PQ} = -0.0168$ for our PQ setup with $N_f = 2$ \cite{PQ} and $f_2 = 
-0.0226$ for $N_f = 2 + 1$ \cite{LR}. 
Thus  the effect of quenching the strange quark is exactly known at NLO: $f_2 - f_2^{PQ} = 
-0.0058$ ($\simeq 26\%$ of $f_2$).
This correction was taken into account in Ref.~\cite{ETMC09} and it has no error.
Note that the difference between the values of $f_2$ at $N_f = 2 + 1$ and $N_f = 2$ is almost 
an order of magnitude less than the difference between those at $N_f = 2 + 1$ and $N_f = 0$.
In our opinion this should be traced back to the facts that $f_2$ is dominated by meson loops
and the pion contribution is the same in the $N_f = 2 $ and $N_f = 2 + 1$ theories.

\begin{figure}[!hbt]

\centerline{\includegraphics[scale=0.8]{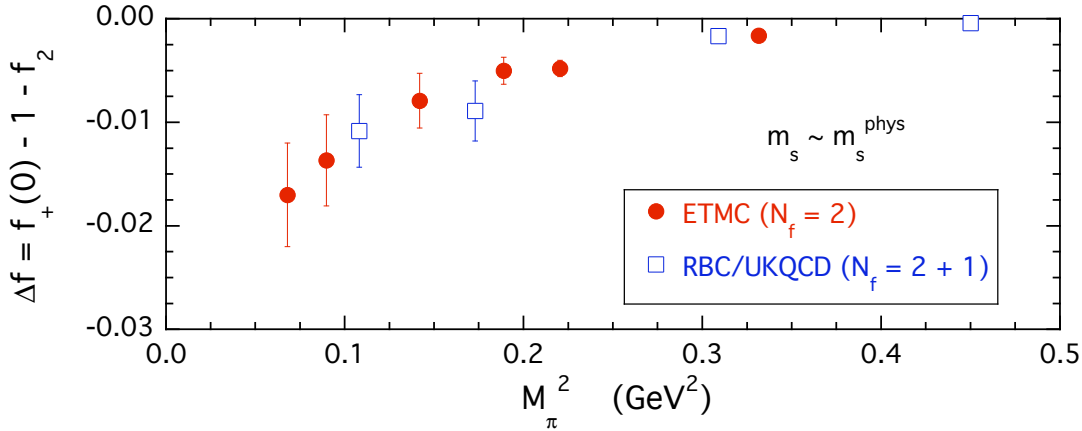}}

\caption{\it Values of the ${\cal{O}}(p^6)$ term $\Delta f$ [Eq.~(\protect\ref{eq:Deltaf})] obtained 
by  the ETM \cite{ETMC09} and RBC/UKQCD \cite{RBC08} Collaborations taking into account 
the values of the NLO term $f_2$ appropriate for $N_f = 2$ and $N_f = 2+1$.}

\label{fig:Deltaf}

\end{figure}

The task is thus reduced to the problem of estimating the quenching effect on the quantity
 \be
    \Delta f \equiv f_4 + f_6 + ... = f_+(0) - (1 + f_2) ~ .
    \label{eq:Deltaf}
 \ee
The results obtained for $\Delta f$ by the ETM Collaboration at $N_f = 2$ \cite{ETMC09} and 
by the RBC/UKQCD one at $N_f = 2 + 1$ \cite{RBC08} are compared in Fig.~\ref{fig:Deltaf}.
It can clearly be seen that the effect of quenching the strange quark is well within the statistical 
uncertainties found by the two Collaborations.

In Ref.~\cite{ETMC09} the relative quenching error on $\Delta f$ has been estimated to be at most 
$50\%$ of the same relative effect on $f_2$.
Such an estimate is based on the observation that, while the NLO term $f_2$ is expected to be 
sensitive to the number of sea-quark flavors being only determined by the contribution of meson 
loops, the ${\cal{O}}(p^6)$ term $\Delta f$ receives important contributions from the local terms 
of the effective theory, which are expected to be dominated by the physics of the nearest 
resonances. 
Our estimate corresponds to a systematic error of $0.0028$ (i.e., $\simeq 13\%$ of $\Delta f$), 
which incidentally turns out to be of the same size of the difference between the ETM result 
for $\Delta f$ at $N_f = 2$ and the quenched one of Ref.~\cite{SPQCDR}.
Thus we expect our estimate of the quenching error to be a quite conservative one.

To close this Section we have collected the budget for the systematic error of the ETM result 
(\ref{eq:final})  in Table \ref{tab:syst}.

\begin{table}[!htb]

\begin{center}
\begin{tabular}{||c||c|c||}
\hline
 $Source$ & $systematic~error$ & $\%~of~[1 - f_+(0)]$ \\ \hline \hline
 $q^2-dependence~and~chiral~extrapolation$  &  $0.0035$ & $~8$ \\ \hline
 $finite~size$                                                              &  $0.0018$ & $~4$ \\ \hline
 $discretization$                                                        &  $0.0037$ & $~8$ \\ \hline
 $quenching~of~the~strange~quark$                  &  $0.0028$ & $~6$ \\ \hline \hline
 $Total~(in~quadrature)$                                        &  $0.0062$ & $14$ \\  \hline \hline

\end{tabular}

\caption{\it Budget of the systematic error for the ETM determination (\protect\ref{eq:final}) of 
the vector form factor at zero-momentum transfer, $f_+(0)$, obtained in Ref.~\cite{ETMC09}.}

\label{tab:syst}

\end{center}

\end{table}

\section{$D \to \pi \ell \nu_\ell$ decays}

In the case of the semileptonic decays of a heavy meson $H$ it is convenient to use a decomposition 
of the matrix element of the weak vector current in which the form factors are independent of the 
heavy-meson mass $M_H$ in the static limit, namely
 \be
    \< P(p_P) | V^\mu | H(p_H) \> = \sqrt{2 M_H} \left[ v^\mu ~ f_v(E) + p_\perp^\mu ~ f_p(E) \right] ~ ,
    \label{eq:HP_static}
 \ee
where $v \equiv p_H / M_H$, $p_\perp \equiv p_P - E v$ and $E \equiv v \cdot p_P = (M_H^2 + 
M_P^2 - q^2) / 2M_H$ is the energy of the final meson in the rest-frame of the initial one.
The relation between the form factors $f_v(E)$ and $f_p(E)$ and those appearing in Eq.~(\ref{eq:HP}) 
is
 \be
      f_+(q^2) & = & \left[ f_v(E) + (M_H - E) f_p(E) \right] / \sqrt{2 M_H} ~ , \\
      f_0(q^2) & = & \left[ (M_H - E) f_v(E) + (E^2 - M_P^2) f_p(E) \right] \sqrt{2M_H} / (M_H^2 - M_P^2) ~ .
      \label{eq:rel}
  \ee

In the static limit both the mass and the energy dependence of the form factors $f_v(E)$ and $f_p(E)$ 
have been investigated  within the Heavy Meson ChPT (HMChPT) in Ref.~\cite{Damir}.
For a pion in the final state, i.e.~$P = \pi$, one has at NLO
 \be
     f_v(E) & = & D_0 \left[ 1 + D_1(E) M_\pi^2 + D_2(E) - 3 (1 + 3 g^2) M_\pi^2 ~ L(M_\pi^2) / 4 \right. \nn \\
                 & - & \left. 2 (E^2 - M_\pi^2) ~ L(M_\pi^2) - 2 M_\pi ~ E ~ F(E / M_\pi) \right] ~ , \\
     f_p(E) & = & C_0 \left[ 1 + C_1(E) M_\pi^2 + C_2(E) - 3 (1 + 3 g^2) M_\pi^2 ~ L(M_\pi^2) / 4 \right] 
                         / (E + \Delta^*) ~ , 
     \label{eq:HMChPT}
 \ee
where $C_i$ and $D_i$ ($i = 0, 1, 2$) are unknown low-energy constants (LEC's), $g$ is the $H^*H\pi$ 
coupling constant (with $H^*$ being the vector resonance of the heavy meson H), $\Delta^* \equiv 
M_{H^*} - M_H$, $L(M_\pi^2) = \mbox{log}(M_\pi^2) / (4 \pi f_\pi)^2$ and $F(x) = 2 \sqrt{x^2 -1} ~  
\mbox{log}[x + \sqrt{x^2 - 1}] / (4 \pi f_\pi)^2$ for $x \geq 1$.

In addition a generalization of the Callan-Treiman relation \cite{Voloshin} constrains the value of 
$f_0(q_{max}^2)$ to be equal to the ratio of the leptonic decay constants $f_H / f_\pi$ at the chiral 
point $M_\pi = 0$.
Therefore, using SU(2) HMChPT for $f_H$ and SU(2) ChPT for $f_\pi$, one gets at NLO
 \be
     f_H / f_\pi = \sqrt{2 / M_H} ~ D_0 \left[ 1 + D_2(0) \right] \left[ 1 + B ~ M_\pi^2 +  (5 - 9 g^2) M_\pi^2  
                          ~ L(M_\pi^2) / 4 \right] ~ , 
     \label{eq:CT}
  \ee
where $B$ is an unknown LEC.
 
We have calculated the vector and scalar form factors $f_+(q^2)$ and $f_0(q^2)$ as well as the 
decay constants $f_H$ and $f_\pi$ \cite{ETMC_fD}, using the gauge configurations generated 
by the ETM Collaboration \cite{ETMC} with $N_f = 2$ flavors of dynamical twisted-mass quarks 
at a single lattice spacing $a \simeq 0.088~\mbox{fm}$ for various values of the sea quark 
mass.
The valence light-quark mass is kept equal to the sea quark mass to get unitary pions with a 
simulated mass ranging from $\simeq 260$ to $\simeq 575~\mev$, as in the study of $K \to \pi 
\ell \nu_\ell$ decays.
For each pion mass we use three values of the charm quark mass to allow for a smooth, local 
interpolation of our results to the physical D-meson mass.
At the two lowest pion masses the lattice volume is $L^3 \cdot T = 32^3 \cdot 64~a^4$, 
while at the higher ones it is $24^3 \cdot 48~a^4$ in order to guarantee that $M_\pi L \gtrsim 3.7$.

We have then applied Eqs.~(\ref{eq:HMChPT}) and (\ref{eq:CT}) for a simultaneous fit of the energy 
and pion mass dependence of our results. 
Simple polynomial parameterizations of the energy dependence of the LEC's $C_{1,2}$ and $D_{1,2}$  
have been adopted, while the value $g = 0.6$ has been taken from Ref.~\cite{Damir_g} and the quantity 
$\Delta^*$ has been fixed at its value at the physical point ($\Delta^* = 138~\mev$) . 
The range of values of $q^2$ covered by our data is quite large, extending from  $q^2 \approx 0$ up 
to $q^2 = q_{max}^2$, which corresponds to values of the energy $E$ up to $\approx 1~\gev$.
The quality of the fit provided by Eqs.~(\ref{eq:HMChPT}) turns out to be remarkably good, though the 
chiral expansion of Ref.~\cite{Damir} is in principle limited to the static limit and to values of the energy 
E well below the scale of chiral-symmetry breaking.

\begin{figure}[!hbt]

\centerline{\includegraphics[scale=0.8]{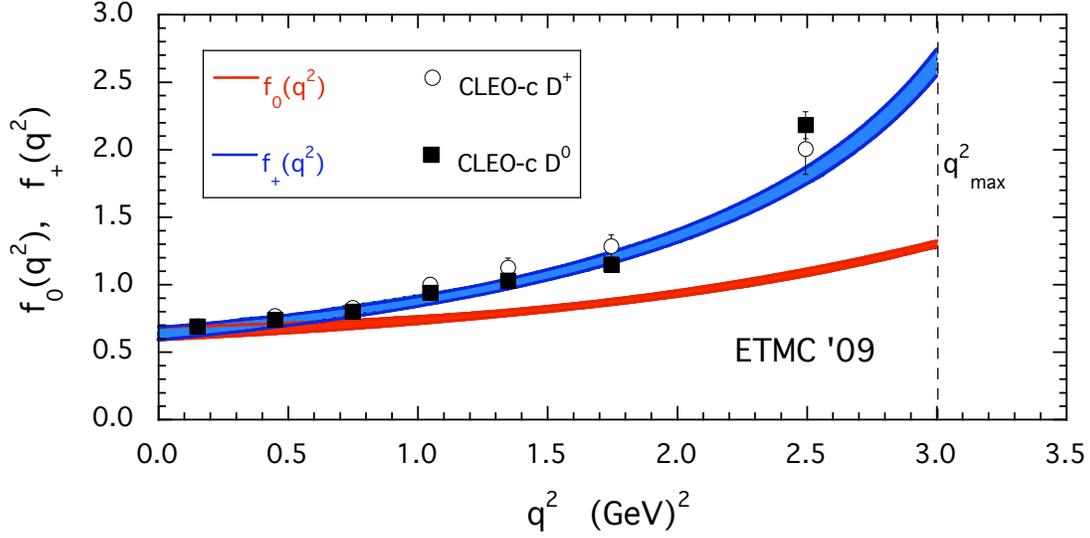}}

\caption{\it Vector [$f_+(q^2)$] and scalar [$f_0(q^2)$] form factors for the $D \to \pi \ell \nu_\ell$ decay 
versus the squared 4-momentum transfer $q^2$. 
The bands correspond to the regions selected at $1 \sigma$ level by the chiral fit (\protect\ref{eq:HMChPT}) 
applied to our lattice results. 
The dots and the squares are the experimental data from Ref.~\cite{CLEOc}.}

\label{fig:DPi}

\end{figure}

The extrapolation of the vector and scalar form factors to the physical pion mass is quite accurate in 
the full range $0 \leq q^2 \leq q_{max}^2$, as shown in Fig.~\ref{fig:DPi}.
Our results are also in good agreement with the latest experimental data from the CLEO Collaboration  
\cite{CLEOc}, obtained after assuming for the CKM matrix element $|V_{cd}|$ the value implied by 
unitarity.
Only around $q^2 \approx q_{max}^2$ our lattice predictions for $f_+(q^2)$ are slightly below the 
experimental results.
At $q^2 = 0$ we get $f_+(0) = 0.64(5)$ which agrees with the lattice result $f_+(0) = 0.64(3)(6)$
obtained in Ref.~\cite{MILC}  with $N_f = 2 + 1$.

\begin{figure}[!hbt]

\centerline{\includegraphics[scale=0.8]{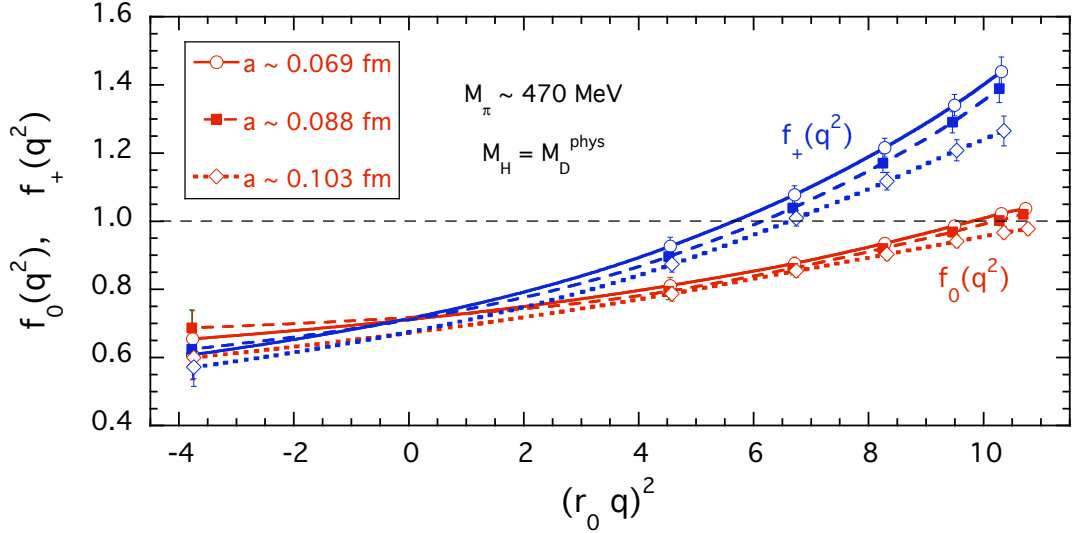}}

\caption{\it Vector (blue markers) and scalar (red markers) form factors for the $D \to \pi \ell \nu_\ell$ decay
versus the squared 4-momentum transfer $q^2$ in units of the Sommer parameter $r_0$, obtained at 
three lattice spacings for $M_\pi \simeq 470~\mev$ and smoothly interpolated at the physical D-meson 
mass.}

\label{fig:DPi_470}

\end{figure}

Since our results have been obtained at a single value of the lattice spacing ($a \simeq 0.088$ fm) and 
a heavy mass, like the charm one, is involved, the question of possible sizable discretization effects 
naturally arises.
We have therefore computed the form factors for $M_\pi \simeq 470~\mev$ at three values of the  lattice 
spacing ($a \simeq 0.069, 0.088$ and $0.103~\mbox{fm}$).
The results are shown in Fig.~\ref{fig:DPi_470}.

It can be seen that discretization effects are small around $q^2 \approx 0$ (at the level of the percent 
between the two finest lattices), while they increase toward $q^2 = q_{max}^2$, particularly in the 
case of the vector form factor $f_+(q^2)$.
Thus we observe that discretization effects related to the presence of the charm quark mass are quite 
limited for our setup.
This may be related to the fact that the form factors are extracted from ratios of correlation functions, 
in which lattice artifacts may partially cancel out.
We observe moreover that  the discretization effects on $f_0(q_{max}^2)$ are of the order of few 
percent, i.e.~similar to those found for $f_D / f_\pi$  in the case of our action \cite{ETMC_fD}.

A complete study of the scaling property of the vector and scalar form factors at various pion masses is 
in progress. 
The results presented in Fig.~\ref{fig:DPi_470}, however, makes us confident that the agreement with 
the experimental data visible in Fig.~\ref{fig:DPi} will not be spoiled by a more detailed analysis of 
discretization effects.

\section*{Acknowledgements}
We thank our ETM collaborators for their help and encouragement. 
This work has been supported in part by the EU ITN contract MRTN-CT-2006-035482, ``FLAVIAnet''.
F.M. also acknowledges the Consolider-Ingenio 2010 Program CPAN (CSD2007-00042) supported 
also by CUR Generalitat de Catalunya under project 2009SGR502.

\end{document}